\documentclass{svproc}
\usepackage[T1]{fontenc}
\usepackage[utf8]{inputenc}

\usepackage{url}
\usepackage{graphicx}

\begin{document}
\mainmatter

\title{Rusty Flying Robots: Learning a Full Robotics Stack with Real-Time Operation on an STM32 Microcontroller in a 9 ECTS MS Course}
\titlerunning{Rusty Flying Robots}  %
\author{Wolfgang Hönig\inst{1,2} \and Christoph Scherer\inst{1,2} \and Khaled Wahba\inst{1}}
\authorrunning{W. Hönig et al.}
\tocauthor{Wolfgang Hönig, Christoph Scherer, Khaled Wahba}
\institute{
Technical University Berlin, Germany\\
\email{\{hoenig, c.scherer, k.wahba\}@tu-berlin.de}
\and
Robotics Institute Germany (RIG)
}

\maketitle

\begin{abstract}
We describe a novel masters-level projects class that teaches robotics along the traditional robotics pipeline (dynamics, state estimation, controls, planning). One key motivational part is that students have to directly apply the algorithms they learn on a highly constrained compute platform, effectively making a robot fly.
We teach nonlinear algorithms as deployed in state-of-the-art flight stacks such as PX4.
Didactically, we rely on two core concepts: 1) avoidance of provided black-box software infrastructure, and 2) usage of the safe and efficient programming language Rust that is used on the PC (for simulation) and an STM32 microcontroller (for robot deployment). We discuss our methodology and the student feedback over two years with ten students each.

Teaching material: \url{https://imrclab.github.io/teaching/flying-robots}.

\end{abstract}

\section{Motivation}

\begin{figure}[tbh]
    \centering
    \includegraphics[width=0.64\linewidth]{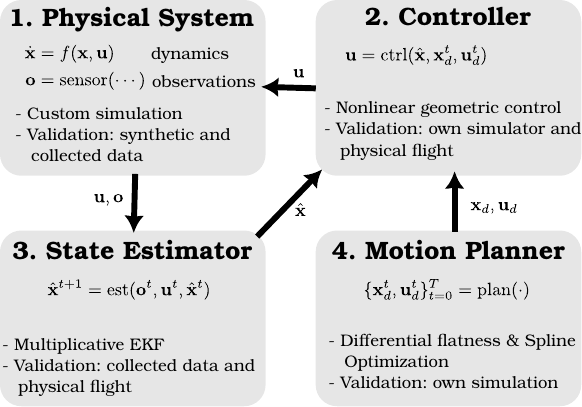}
    \includegraphics[width=0.34\linewidth]{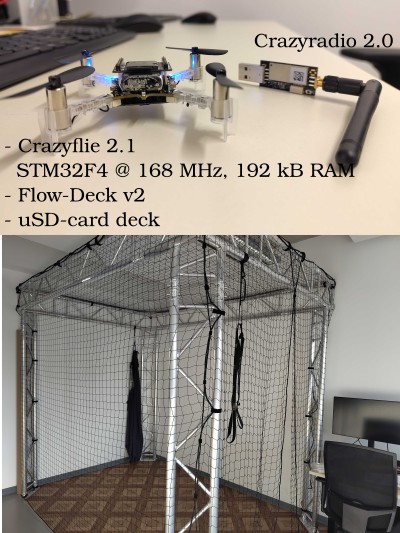}
    
    \caption{Overview of the class. There are four major parts, where students learn about mathematical foundations and implementing their own major parts of a modern flight stack. Code is written in embedded Rust and has to be successfully deployed on an STM32 microcontroller, computing control actions and state estimates in real-time.}
    \label{fig:overview}
\end{figure}

Deploying algorithms on robots is challenging, because they need to run in real-time to be able to react to new observations.
Practically, companies often spend a lot of engineering effort to implement algorithms in low-level languages such as C for code that runs on microcontrollers, or C++ for code that runs on small-board computers.
Such time effort is typically not suitable for classes or projects at universities.
From an educational perspective this creates a dilemma: one can focus on very simplistic algorithms that can be implemented in a reasonable time in a low-level language to demonstrate that it runs on hardware, focus on more advanced settings and provide sophisticated software infrastructure (e.g., ROS 2~\cite{ROS2} packages, robotics-specific frameworks like Drake~\cite{drake}), or focus on non-realtime scenarios in simulation (e.g., Python that operates on datasets or interacts with a simulator such as Gazebo).
All three options have the downside that they reduce the appeal of using robotics to foster interest and excitement in \emph{science, technology, engineering, and mathematics} (STEM) education: when learning about science and technology, one can solve challenging tasks in the real world.
We created a practical 9 ECTS Master-level course that demonstrates the full robotics pipeline on the example of making a commercially-off-the-shelf multirotor fly with state-of-the-art nonlinear algorithms (see Fig.~\ref{fig:overview}).
Students learn about the mathematical foundations and have to implement a simulator, controller, state estimator, and motion planner such that operation in real-time is possible on an STM32 microcontroller (168 MHz, 192 kb RAM).
 
Rather than using the traditional approach of implementing in Python and/or C, we adopt Rust, a more recent programming language that has advantages over C(++) and Python for robotics.
Rust is a compiled language with full memory management control (similar to C/C++), a borrow checker that can provide safety guarantees at compile-time (unlike C/C++), and support for embedded microcontroller target architectures (unlike Python).
Syntactically, it is related to C(++) and it has features like generics and operator overloading that allow to write type-safe mathematical equations (unlike C).
We found that even though the majority of students did not have any prior experience with Rust, the overarching goal of implementing state-of-the-art nonlinear components of a modern robotics pipeline can be achieved within a 9 ECTS class, without providing any additional software components such as math libraries.

\section{Related Educational Concepts}

Teaching the full robotics stack (dynamics, decision making, state estimation, controls) as a system architecture is very common in robotics education, in particular on the undergraduate level~\cite{welton_modular_1993,jung_experiences_2012,ikeda_guiding_2023}.
Graduate level courses can then dive into a particular topic (e.g., state estimation) to provide a deeper theoretical understanding.
Previous studies have shown that in all cases \emph{active learning}, an approach where students are actively involved in the learning process, is effective in robotics~\cite{lopeznicolas_active_2014,chan_addressing_2018,lobo_preparing_2021,yang_robotics_2024}.
One special case is project-based learning~\cite{cappelleri_robotic_2013,tosello_training_2016,farzan_project-based_2023}, which is often applied for graduate-level training to ensure that students can develop solutions for real robotic problems.
For general engineering degrees it might be beneficial to combine mechanics, controls, electronics, and software in a curriculum~\cite{berry_practical_2020}.
Our course uses active learning by letting students experience building a simulator as well as all needed software components to deploy sophisticated, graduate-level algorithms on highly constrained embedded hardware.

Embedded programming and robotics can be taught in a synergetic fashion~\cite{hamrita_robotics_2005,braunl_robotics_2007,bruce_teaching_2013,lee_development_2021}, where a motivational factor is that written code is directly deployed on hardware.
Given that using simulation tools before deployment is crucial in practice, many educational concepts rely on simulators~\cite{camargo_systematic_2021,cruz-ramirez_nao_2022,yindeemak_robotic_2025}.
However, curricula in robotics do not routinely include the knowledge on how to build appropriate simulation tools.
Recently, browser-based tools may guide students in their experiments~\cite{stein2025novice} (here, one written in Rust).
Our course combines simulation and real-robot deployment using a single programming language.

Courses covering flying robots have been around for many years~\cite{gaponov_quadcopter_2012,bolam_curriculum_2017,canas_open-source_2020,carlone_visual_2022,he_introductory_2023}, including some that are more aeronautics-focused~\cite{bolam_curriculum_2017,lobo_preparing_2021,he_introductory_2023}.
The most similar to our course have a robotics-focus and are described in prior papers~\cite{gaponov_quadcopter_2012,canas_open-source_2020,carlone_visual_2022}, were previously offered on Coursera (Aerial Robotics), or are summarized in a text book~\cite{2025-mueller-DynamicsControlAutonomous}.
In contrast to prior classes, we uniquely combine implementing state-of-the-art algorithms with deploying on low-power microcontrollers.

\section{Methodology}

The class is designed as a practical 9 ECTS (270 h) course.
A small fraction is allocated for lectures (around 10 h) and discussions (around 20 h), while the rest is hands-on pratical work both in simulation and on a real robot.
As platform, we use a Bitcraze Crazyflie 2.1~\cite{bitcraze} commercially-off-the-shelf (CoTS) open-source and open-hardware quadrotor with two CoTS extensions (for optical flow and data recording).
This robot is safe to use for humans due to its low weight (around 35 g with added sensors) and no special external equipment is needed.
However, we do operate it in a room with padded floor and safety nets to reduce hardware failures and recommend the use of safety googles.

The class is structured in four major parts (see subsections below) following a standard robotics pipeline.
The course starts with introductory material to flying robots (history, kinds of robots, applications, example videos), in part inspired by~\cite[Chapter 52]{springerHandbook}.
The discussion session allow a structured exchange with and between students, where we openly discuss challenges, obtained results, and expectations.
Each assignment is graded on the functionality, code quality, and ability for students to answer questions about their code.
Tools, including generative AI, are allowed but students have to understand the code in order to successfully pass the oral discussions.

\subsection{Simulation (4 weeks)}

Many roboticists use simulation tools (e.g., Gazebo, CoppeliaSim, MuJoco, WeBots, or Isaac Sim) in their everyday work, for example to validate algorithms, tune parameters, or train policies.
These tools are most beneficial when the students first understand first the underlying physics and dynamics.  
Thus, this course forces the students to write their own multirotor simulation first.
The positive side effect is that students learn about numerical challenges as well as the foundational mathematical operations in SE(3).
For didactical reasons, we first introduce the 1D case (the robots' state is its height and vertical velocity; the action is the upwards force that a thruster generates to compensate gravity) followed by the 2D case (the robot has two thrusters and operates in the $y-z$ plane).
For the 1D case, we also provide and discuss example code of a basic simulator (37 lines of Rust code, no additional libraries needed) and an example for visualization of the resulting motion in the webbrowser using \texttt{MeshCat}~\cite{meshcat}.

The first assignment is to implement a simulator for the 3D case in Rust.
In the lecture, we introduce the relevant mathematical operations, focusing on quaternions for operations in SO(3).
Students have to implement their own math library with all needed operations (such as integrating the rotation given the angular velocity), for easy transfer to the microcontroller later on (where std and heap allocations should be avoided).
The simulator has to be validated rigorously by integrating actions using a) given trajectories that were derived mathematically using differential flatness for known parameters, and b) data collected directly from the robot.
A successful validation in b) requires (basic) system identification of key parameters of the provided platforms by the students.
We found that proper validation of the simulation is crucial, as errors in later assignments may compound otherwise.

\subsection{Controls (4 weeks)}

A particular control challenge for robotics is the operation on the manifold that is induced by the operation in SE(3).
Moreover, the dynamics of the system are nonlinear, such that classic techniques such as LQR controllers have a limited applicability.
We consider a state-of-the-art geometric controller that has exponential stability guarantees~\cite{lee2010geometric} and where variants of this controller can be found in popular flight control stacks such as PX4~\cite{px4}.
In the lecture we introduce the control law mathematically (about 10 equations that mostly rely on linear algebra and use rotation matrices internally).

The students implement and validate their controller in their own simulation, demonstrating that given reference trajectories can be safely followed.
Using their own simulator (rather than providing a simulation tool) has the advantage that students can easily debug problems directly and also have control over all physical parameters (e.g., inertia matrix or motor constants), that are otherwise often not directly adjustable.
Once the simulation works, they can integrate their code into the open-source STM32 flight software.
This firmware is written in C and has one example on how to integrate a Rust library (that essentially prints a string).
Students learn about interoperability of C and Rust and relevant software tools (e.g., \texttt{bindgen}, which can generate Rust bindings to existing C interfaces). 
They have to convert the data types (e.g., states and actions), test and debug their own controller.
This is challenging, as small mistakes (or poorly tuned variables) lead to crashes of the robot within a few milliseconds.
The transfer from simulation to the robot is often quickly successful; issues mostly arise from the language transition C/Rust as well as unit conversions.
The instructors know common pitfalls and guide the students during the sim-to-real transfer in a hands-on fashion.
The hardware is robust enough to sustain crashes without the need of expensive repairs.

\subsection{State Estimation (3 weeks)}

The goal of state estimation is to use the on-board sensors (gyroscope, accelerometer, optical flow) to estimate the robots' state (position, orientation, velocity).
The operation on the SO(3) manifold makes it non-trivial to apply vanilla versions of standard approaches such as the Extended Kalman Filter (EKF).
The lecture introduces state-of-the-art SO(3) complementary filters~\cite{2008-mahony-NonlinearComplementaryFilters,2014-madgwick-AHRSAlgorithmsCalibration} as well as the multiplicative EKF (MEKF) variant that can directly handle SO(3)~\cite{2003-markley-AttitudeErrorRepresentations}.
The MEKF is mathematically simple (4 equations with linear algebra operations), however practically very difficult to implement to ensure numeric stability and operation in real-time.
This is due to the need of a matrix inversion of a matrix that is $9\times 9$ for our 9-dimensional state-space.
We introduce various practical approaches to tackle these challenges: symbolic differentiation and simplifications using \texttt{sympy}~\cite{sympy} (full examples are provided), and a mathematical trick known as scalar updates.

Students are tasked to implement the filter in simulation first.
Here, a dataset collected with the real robot is given as well as reference state estimates from the existing on-board MEKF.
Once the filter works in simulation, it has to be ported to the STM32 microcontroller and its performance demonstrated on real flights.
This assignment often forces students to significantly improve their math library to use generics. Given the higher computational burden compared to the non-linear controller, it may also require time to improve the overall runtime of the written code. Students have the option to implement the much simpler complementary filters only, although this comes with a point deduction.

\subsection{Planning (3 weeks)}

For decision making, we focus on motion planning of agile maneuvers through narrow passages.
Like many other mobile robots, multirotors are differentially-flat.
This property allows to plan smooth trajectories of the robot's three-dimensional position and compute all other relevant variables (including nominal controls) thereafter.
In the lecture, we provide a mathematical description and hints for their derivation.
Afterwards, we introduce how smooth positional trajectories can be efficiently computed using quadratic programs (QPs).
QPs can be solved in real-time within a few milliseconds on highly constrained microcontrollers, using designated solver generators~\cite{cvxgen,osqp-codegen,nguyen2024tinympc}.
For the last assignment, students have to a) implement a motion planner by defining the QP and using an existing solver, and b) implement the differential flatness to extract reference states and nominal controls.
The resulting aggressive maneuvers have to be validated in their own simulator.

\section{Results}

The class was conducted in fall 24/25 and fall 25/26 and each time limited to 10 students due to its practical nature with occasional 1-on-1 coaching.
Towards the end of the class, an official survey was conducted by the university (participation: 8+8 students).
Surveyed students studied Computer Science, Computer Engineering, or Electrical Engineering with the majority in the third and fourth semester.
Half of the students reported to need 6-9 h/week, one quarter to need 12-16 h/week, and another quarter only listed 2-4 h/week, keeping the overall effort well within the expected effort for a 9 ECTS class.
At the same time, the majority of the students had no prior experience in Rust (but experience usually in both C and Python) and had to learn a new programming language.
For this, students had to work auto-didactically as the lecture only included a high-level overview and a few examples of Rust.
Overall, the class was rated as very good (mean of 1.1, where 1 is the best and 5 the worst).
The most positive comment was ``I really enjoyed and learned a lot from this lecture. Was genuinely one of the best yet.'' and ``Such a fun lecture, I really enjoy it despite needing quite a lot of time for the implementations''.
The most negative comments suggested improvements in making the lecture material more comprehensive (e.g. by having a script) and to split up some of the assignments into smaller deliverables.

\section{Conclusion}

Teaching a robotics class with Rust as a programming language has many advantages: compiled code tends to work and executes efficiently (unlike using Python only), deployment even on highly-constrained embedded systems is possible without rewriting (unlike using Python followed by conversion to C), and the code is compact even for vector and matrix operations (unlike C only).
The obvious expected downside of students having to learn yet another language seems to be a non-issue and may act as a motivating challenge instead.

Our work provides a blueprint on how to effectively teach foundations of flying robots that can be readily deployed in other universities thanks to the openly accessible material.
At the same time we believe that some of the core concepts can be used beneficially for other robotics classes, namely the philosophy of less reliance on pre-defined software infrastructure (such as simulators or math libraries) and the idea of using Rust both as a high-level and low-level language for code that needs to run on the hardware.
In the future, we are planning to investigate if similar approaches can benefit undergraduate robotics education.

\bibliographystyle{spmpsci}  %
\bibliography{references,class_refs,webpages}    %

\end{document}